\documentclass[a4paper,twocolumn,prl,showpacs,showkeywords,aps,nofootinbib]{revtex4}
\usepackage{graphicx}
\usepackage{amsmath}
\usepackage{bm}
\newcommand{\beq}{\begin{equation}}
\newcommand{\eeq}{\end{equation}}
\newcommand{\beqa}{\begin{eqnarray}}
\newcommand{\eeqa}{\end{eqnarray}}
\def\ra{\rangle}
\def\la{\langle}

\newcommand {\fexp} [1] {\exp \left[ #1 \right]}

\newcommand {\ms} {\, \mbox{ms}}

\begin{document}
\title{Fast optimal frictionless atom cooling in harmonic traps}

\author{Xi Chen$^{1,2}$}

\author{A. Ruschhaupt$^{3}$}
\author{S. Schmidt$^{3}$}
\author{A. del Campo$^{4,5}$}
\author{D. Gu\'ery-Odelin$^{6}$}
\author{J. G. Muga$^{1}$}
\affiliation{$^{1}$ Departamento de Qu\'{\i}mica-F\'{\i}sica,
UPV-EHU, Apdo 644, 48080 Bilbao, Spain}

\affiliation{$^{2}$ Department of Physics, Shanghai University,
200444 Shanghai, P. R. China}

\affiliation{$^{3}$ Institut f\"ur Theoretische Physik, Leibniz
Universit\"{a}t Hannover, Appelstra$\beta$e 2, 30167 Hannover,
Germany}

\affiliation{$^{4}$Institute for Mathematical Sciences,
Imperial College London, 53 Prince's Gate, SW7 2PG London, UK}
 
\affiliation{$^{5}$QOLS, Blackett Laboratory, Imperial College London,
Prince Consort Road, SW7 2BW London, UK}

\affiliation{$^{6}$Laboratoire Collisions Agr\'egats R\'eactivit\'e, CNRS UMR 5589, IRSAMC, Universit\'e Paul Sabatier, 118 Route de Narbonne, 31062 Toulouse CEDEX 4, France}

\begin{abstract}
A method is proposed to cool down atoms in a harmonic trap without phase-space compression 
as in a perfectly slow adiabatic expansion, i.e., keeping the populations of the instantaneous initial and final levels invariant, but in a much shorter time. This may require that the harmonic trap becomes an expulsive parabolic potential
in some time interval. The cooling times achieved are also shorter than previous minimal times using optimal-control bang-bang methods and real frequencies.      
\end{abstract} 


\maketitle

A fast adiabatic expansion in a short finite time 
looks like a contradiction in terms. An ``adiabatic'' process in
quantum mechanics is a slow process where the system {follows} at all times the 
instantaneous eigenvalues and eigenstates
of the time-dependent Hamiltonian. This is in a sense maximally efficient as 
the populations do not change, i.e. there is no heating or friction, but the price to pay is that the long times needed may render the process useless or 
even impossible to implement.       
Thus, a highly desirable goal is   
to prepare the same final states and energies  
of the adiabatic process in a given finite time $t_f$, without necessarily following the instantaneous eigenstates along the way. We would also like the process to be robust with respect to arbitrary initial states. If fulfilled, this old goal \cite{LR69} has important implications. In particular, cooling
without phase-space compression, which is all that is needed 
for many applications other than Bose Einstein condensation,
could be performed in fast cycles increasing, for example, 
the flux of cold atoms produced and the signal to noise ratio in an atomic
clock \cite{Bize}, or in cold-atom pulsed beam experiments and related technology
\cite{pulsed}.    
This goal also includes as a particular case a long standing question 
in the fields of optimal control theory and finite time thermodynamics, namely,
to optimize the passage between two thermal states of a system \cite{Rabitz,Kos2000,K09,K09b}.  
For time-dependent harmonic oscillators, minimal times have been established 
using ``bang-bang'' real-frequency processes believed up to now to be optimal \cite{K09}, in which the
frequencies are changed suddenly at certain instants but kept constant otherwise.   
In this letter we shall describe a robust solution to the stated general goal for
atoms trapped in a time-dependent harmonic oscillator which applies both to equilibrium and non-equilibrium states. In particular we describe 
cooling processes performed in a time interval smaller than the minimal time
of the bang-bang methods considered so far. 
We shall for simplicity describe our method for states representing single atoms of mass $m$, but the same results are immediately applicable to $N$-body non-interacting fermions or to a Tonks-Girardeau gas \cite{MG}, and generalizations will be relevant for other driving processes, such as cold atom launching or the transport of ultracold atoms with optical tweezers \cite{David2}.        
 
We consider an effectively one dimensional time dependent harmonic oscillator, $H=\hat{p}^2/2m+m\omega(t)^2\hat{q}^2/2$, with an initial angular frequency $\omega(0)>0$ at time $t=0$ and final frequency $\omega_f=\omega(t_f)<\omega(0)$
at time $t_f$. (This amounts to a temperature reduction by a factor 
$\omega_f/\omega(0)$ if the initial and final states are canonical.) The challenge is to find a trajectory 
$\omega(t)$ between these two values so that the populations of the oscillator levels $n=0,1,2...$ at $t_f$ are equal than the ones at $t=0$.   
Our main tool to engineer $\omega(t)$ and the state dynamics will be the  solution of the corresponding Schr\"odinger equation based on the existence of invariants of motion \cite{LR69,Berry,Dodonov,Lohe} of the form $I(t)=1/2[(1/b^2) \hat{q}^2m \omega_0^2+\frac{1}{m}
\hat{\pi}^2]$, where $\hat{\pi}=b(t)\hat{p}-m\dot{b}\hat{q}$ plays the role of a momentum conjugate to $\hat{q}/b$, the dots are derivatives with respect to time, and $\omega_0$ is in principle an arbitrary constant. The scaling, dimensionless function $b=b(t)$ satisfies the subsidiary condition 
\beq\label{subsi}
\ddot{b}+\omega(t)^2 {b}=\omega_0^2/b^3,
\eeq
an Ermakov equation where real solutions must be chosen to make $I$ Hermitian \cite{Erma}. 
$\omega_0$ is frequently rescaled to unity by a scale transformation of $b$ \cite{LR69}.
Other convenient choice is $\omega_0=\omega(0)$ as we shall see. 
$I(t)$ has the structure of a harmonic oscillator Hamiltonian as well (as long as $\omega_0^2>0$),  
with time-dependent eigenvectors $|n(t)\ra$ and time-independent eigenvalues
$(n+1/2)\hbar\omega_0$. The general solution of the Schr\"odinger equation 
is a superposition of orthonormal ``expanding modes'' $\psi(t,x)=\sum_n c_n e^{i\alpha_n(t)}\la x|n(t)\ra$ where 
$\alpha_n(t)=-(n+1/2)\omega_0\int_0^t dt'/b^2$, and the $c_n$ are time independent amplitudes.   
For a single mode and $\omega_0^2>0$, 
\begin{widetext}
\begin{eqnarray*}
\Psi_n (t,x) = \left(\frac{m\omega_0}{\pi\hbar}\right)^{1/4} 
\!\frac{1}{(2^n n! b)^{1/2}}
\fexp{-i (n+1/2) \int_0^t dt'\, \frac{\omega_0}{b(t')^2}}
\fexp{i \frac{m}{2\hbar}\left(\frac{\dot{b}}{b(t)} +
 \frac{i\omega_0}{b^2}\right)x^2}
H_n\left[\left(\frac{m\omega_0}{\hbar}\right)^{1/2}\frac{x}{b}\right],
\label{emode}
\end{eqnarray*}
\end{widetext}
with time dependent average energy   
\beq
\la H(t)\ra_n=\frac{(2n+1)\hbar}{4\omega_0}
\left(\dot{b}^2+\omega(t)^2b^2+\frac{\omega_0^2}{b^2}\right). 
\label{ener}
\eeq
%
The average position is zero and the standard deviation
$\sigma=(\int dx x^2 |\Psi_n|^2)^{1/2}$ 
is proportional to $b$,  
$\sigma=b(n+1/2)^{1/2}/(m\omega_0/\hbar)^{1/2}$, which underlines the  
physical meaning of the scaling factor.   
 
A much studied case corresponds to 
the frequency scaling $\omega(t)=\omega(0)/b^2$ 
with $b=(At^2+2Bt+C)^{1/2}$ \cite{Berry,Dodonov,Ciftja}.
Substituting this in the subsidiary condition gives $\omega_0^2=\omega(0)^2+AC-B^2$.
For a trap with hard walls, the square-root-in-time scaling factor $b$ ($A=0$) has been shown to provide fast and efficient cooling \cite{Schmidt,st}. However, for harmonic traps, much more commonly realized in ultracold experiments, 
such time dependence leads to negative values of $\omega_0^2$ even for modest cooling objectives. This makes Eq. (\ref{emode}) invalid and,
moreover, linear combinations of a continuum of non-square-integrable  expanding modes would be needed to describe the evolution of any single eigenstate of the initial trap. 
This is of course only a drawback to calculate the dynamics, not to  realize the expansion in the laboratory.
Numerical results using other (adiabatic basis) methods \cite{Rice} show that, even though the root-in-time scaling is singularly efficient for adiabatic following as discussed below, the cooling performance fails 
for very short expansion times $t_f$.  An 
alternative, successful strategy put forward here, inspired in inverse scattering techniques for complex potential optimization \cite{Brouard,Palao,Rus}, is to leave $\omega(t)$ undetermined at first and 
impose properties on $b$ and its derivatives at the boundaries, $t=0$ and $t_f$, 
to assure: (a) that any eigenstate of $H(0)$ evolves as a single expanding mode and that (b) this expanding mode becomes, up to a position-independent phase factor, equal to the corresponding eigenstate of the Hamiltonian $H(t_f)$
of the final trap. This keeps the populations in the instantaneous basis equal at the initial and final times.  After $b(t)$ and its derivatives are fixed at the boundaries, $b(t)$ may be chosen as a real function satisfying the boundary conditions, for example as a polynomial or some other convenient functional form 
with enough free parameters.  
Once $b(t)$ has been determined, the physical frequency $\omega(t)$ is 
obtained from the subsidiary equation (\ref{subsi}).    

Let us first discuss the conditions at $t=0$. By choosing $b(0)=1$, $\dot{b}(0)=0$,
$H(0)$ and $I(0)$ commute and have common eigenfunctions
at that instant. We set $\omega_0=\omega(0)$ from now on 
so that $\ddot{b}(0)=0$ must hold as well. These boundary conditions 
imply that any initial eigenstate of $H(0)$, $u_n(0)$,  
will evolve according to the expanding mode (\ref{emode})
for all later times.
In general $H(t)$ and $I(t)$ will not commute for $t>0$, so that the  expanding mode $\Psi_n(t)$ may have more than one component in the ``adiabatic basis'' of instantaneous eigenstates of $H(t)$, $\{u_n(t)\}, n=0,1,2...$, where $u_n(t)=\left(\frac{m\omega(t)}{\pi\hbar}\right)^{1/4} 
\!\frac{1}{(2^n n!)^{1/2}}
\fexp{-\frac{m}{2\hbar}{\omega(t)}x^2}
H_n\left[\sqrt{\frac{m\omega(t)}{\hbar}}{x}\right]$.     
At time $t_f$ we want $\Psi_n(t_f)$ to be proportional, 
up to the global phase factor $e^{i\alpha_n(t_f)}$,
to the corresponding eigenstate of the final trap $u_n(t_f)$.
To this end we impose 
$b(t_f)=\gamma=[\omega_0/\omega_f]^{1/2}$,
$\dot{b}(t_f)=0$, $\ddot{b}(t_f)=0$. 
From Eq. (\ref{ener}),  
%
one finds $\la H(t_f)\ra_n$ in terms of $b_f = b(t_f )$
and $\dot{b}_f = db(t)/dt|_{t=t_f}$.
Since $b_f$ and $\dot{b_f}$ can be set independently we can minimize 
the terms depending on them separately, and    
the global minimum is found to be precisely at the adiabatic energy
$(n+1/2)\hbar\omega_f$, which corresponds to our boundary conditions.  
Any other choice would necessarily produce ``frictional heating''. 
     
\begin{figure}[t]
\begin{center}
\includegraphics[width=0.6\linewidth]{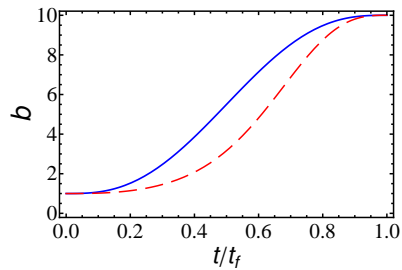}
\end{center}
\caption{\label{fig1} (color online). Examples of ansatz for $b$:  A simple polynomial ansatz (solid line, Eq. (\ref{ans})),
and an exponential of a polynomial (dashed line, $\exp{\sum_{j=0}^5 d_j t^j}$).   $\omega(0) = 250\times2\pi$ Hz,
$\omega(t_f) = 2.5\times2\pi$ Hz, $\gamma =10$.}
\end{figure}

Substituting the simple polynomial ansatz   
\beq\label{ans}
b(t)=\sum_{j=0}^5 a_j t^j
\eeq
into the six boundary conditions 
gives six equations that can be solved to provide the coefficients,   
%
$
b (t) =
6 \left(\gamma -1\right) s^5
-15 \left(\gamma-1\right) s^4 +10 \left(\gamma-1\right)s^3
+ 1,
$
%
where $s=t/t_f$, see Fig. \ref{fig1}. 
%
%
%
%
At initial and final times $0$ and $t_f$,  $\omega(t)=\omega_0/b(t)^2$,  
but this relation 
does not hold in general for an arbitrary intermediate time.      
%
%
%

The above mentioned six conditions leave time dependent phases $e^{i\alpha_n(t)}$ of no relevance regarding the population of the $n$th level. In particular stationary density operators with respect to $H(0)$   (e.g. a canonical state, or a pure state $|u_n(0)\ra\la u_n(0)|$) are mapped onto the corresponding stationary states of $H(t_f)$ with the phases canceled. 
In other cases the phases remain but the populations are preserved. 
Note that $e^{i\alpha_n(t)}$, see Eq. (\ref{emode}), is the phase factor that the initial state $u_n(0)$ would acquire in a virtual adiabatic process in which the adiabatic (instantaneous) energy had the form $(n+1/2)\hbar\omega_0/b^2$.  
Phase control may as well be imposed by adding integral conditions such as  
%
$\tau(t_f) = \int_0^{t_f} dt\, \frac{1}{b(t)^2} = \frac{\omega_f}{\omega_0} t'$,
%
where $t'$ is some desired time. This of course requires   
an ansatz more complicated than Eq. (\ref{ans}), such as a polynomial of higher degree.
 
Numerical examples of frequencies $\omega(t)$ and energies $\langle H(t)\rangle$ of fast adiabatic-like expansions are provided in Figs. 2-4  
using the $b$ shown in Fig. \ref{fig1} for $\omega_0 = 250\times2\pi$ Hz and $\omega_f = 2.5\times 2\pi$ Hz ($\gamma =10$).
These values can be found 
in actual experiments \cite{exp}. 
We could formally study sub-hertz frequencies $\omega_f$
but they would render the trap very sensitive to low-frequency acoustic noise \cite{David}. Compare first the finite times considered 
(from $2$ to $25 \ms$) with the times necessary for actual adiabatic following 
during the whole interval $0<t<t_f$. The adiabaticity condition for the harmonic oscillator becomes 
$|\sqrt{2}\dot{\omega}/(8 \omega^2)|\ll 1$. For a linear ramp, 
$\omega(t)\to \omega_0+(\omega_f-\omega_0)t/t_f$, this implies a very long time, $t_f\gg1.1$ s. In fact it would be  necessary to expand the trap for 6 s to achieve a 1$\%$ relative error in the final energy of the ground state. 
A much more efficient strategy is to distribute $\dot{\omega}/\omega$ uniformly along the trajectory, 
i.e., $\dot{\omega}/\omega^2=c$, $c$ being constant. Thus the trap expansion speed decreases with the splitting. Solving this differential equation and imposing $\omega_f=\omega(t_f)$ we get $\omega(t)=\omega_0/[1-(\omega_f-\omega_0)t/(t_f\omega_f)]$. This corresponds to the case $A=0$, $2B=-(\omega_f-\omega_0)/(t_f\omega_f), C=1$ (i.e., a square-root-in-time scaling factor),
and implies $t_f\gg 11 \ms$ from the adiabaticity condition. A 1$\%$ error level for the ground state energy is achieved after $45 \ms$.      

\begin{figure}[t]
\begin{center}
%
\includegraphics[width=0.46\linewidth]{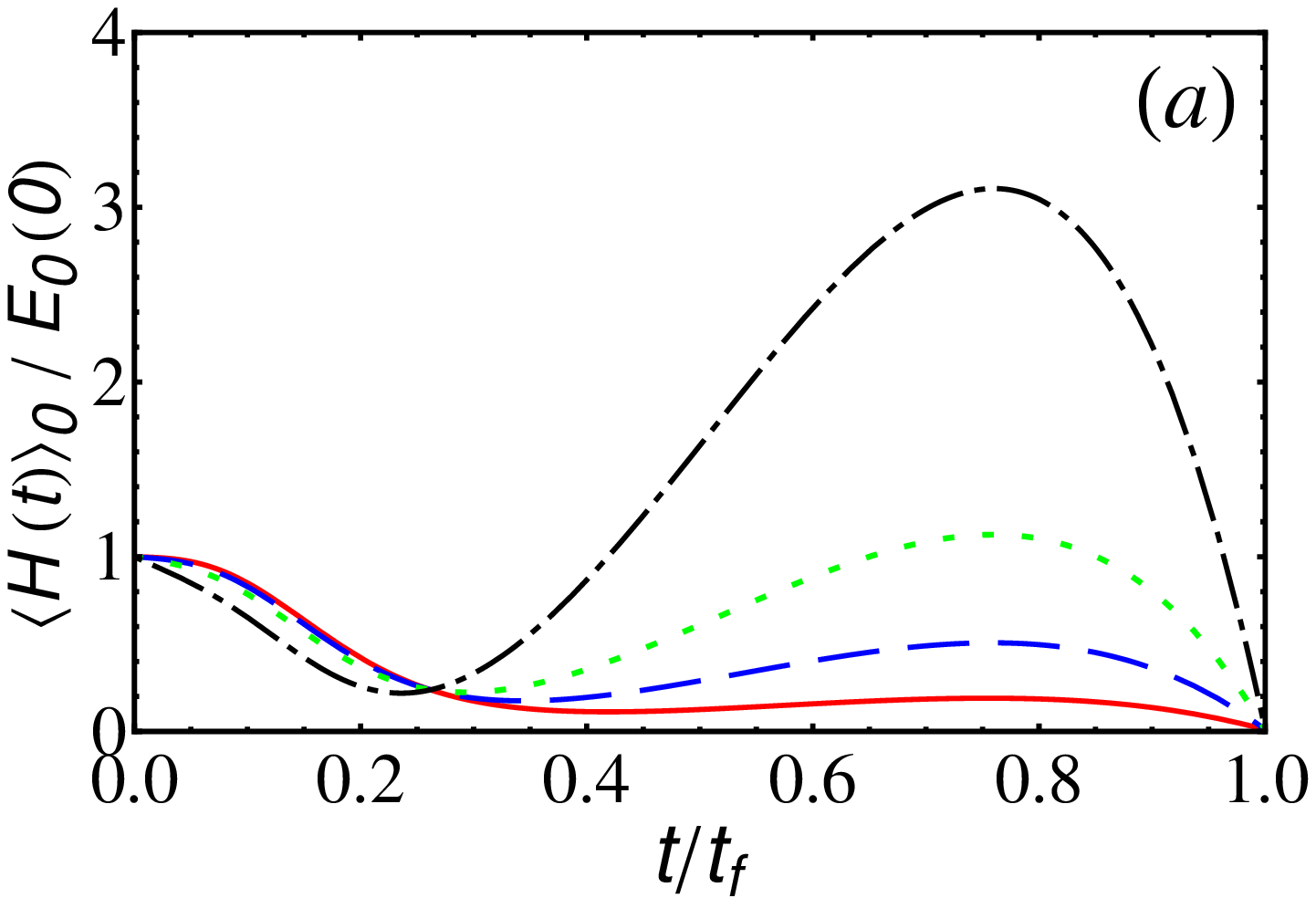}
\includegraphics[width=0.49\linewidth]{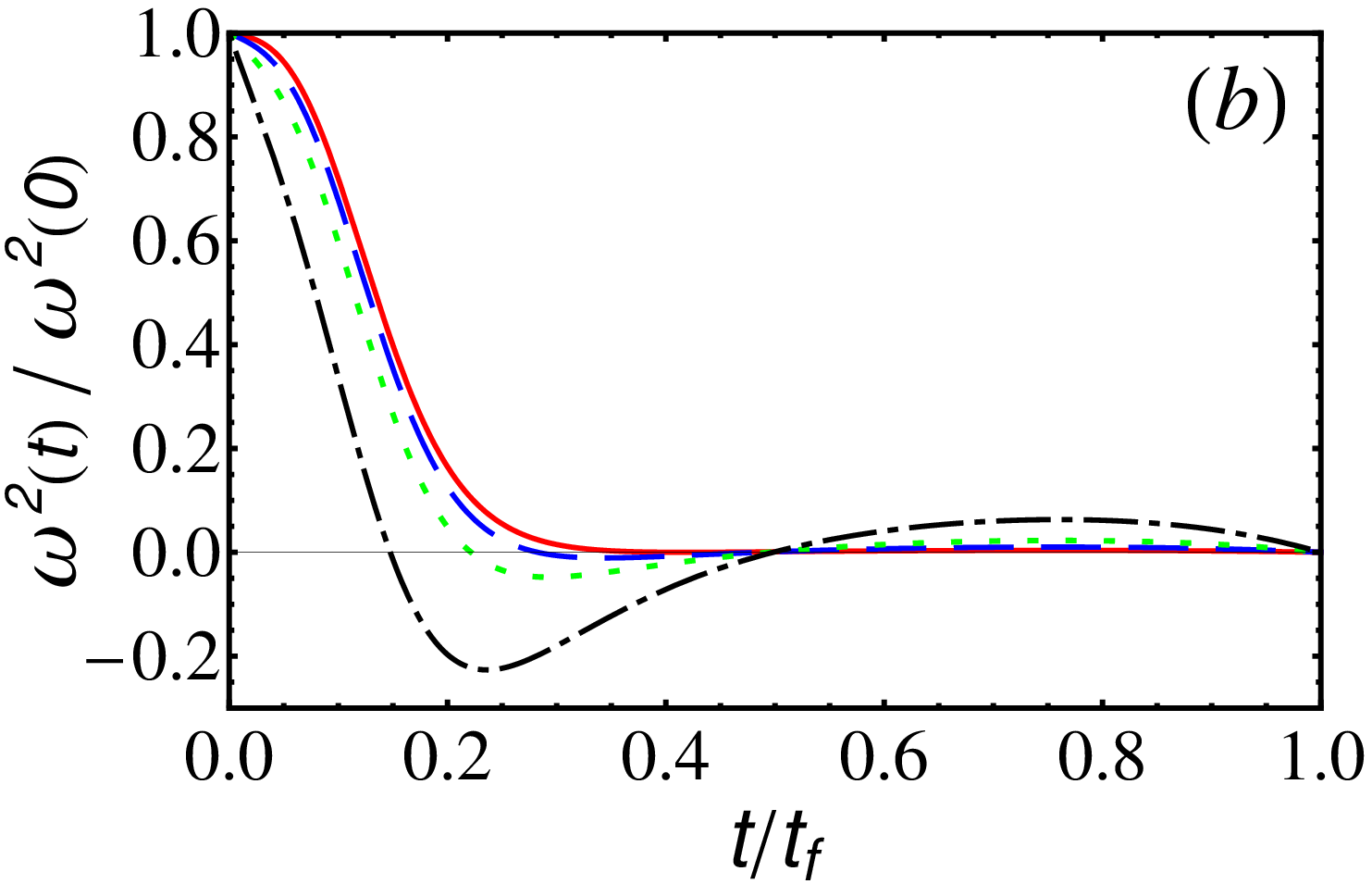}
\end{center}
\caption{\label{fig2}(color online). (a) The average energies of the ground state
expanding mode
for different final times $t_f$: $t_f = 25 \ms$ (solid), $t_f = 15 \ms$ (dashed), $t_f = 10 \ms$ (dotted), and $t_f = 6 \ms$ (dash-dotted). Other parameters as in Fig. \ref{fig1} (polynomial $b$) 
(b) The corresponding squared frequency $\omega(t)^2$.}
\end{figure}

A prominent feature of the trajectories, see Fig. 2b, is that $\omega(t)^2$ may be negative
during some time interval in which  
the potential becomes an expulsive parabola. 
This is physically feasible and has been realized experimentally
with an offset magnetic field that overcomes the optical dipole well 
in the axial direction of elongated cigar-shaped optical traps \cite{Salomon}.   
In general the (imaginary) frequency of the repulsive region increases for shorter cooling times as shown in Fig. \ref{fig2}b.  

\begin{figure}[t]
\begin{center}
%
\includegraphics[width=0.46\linewidth]{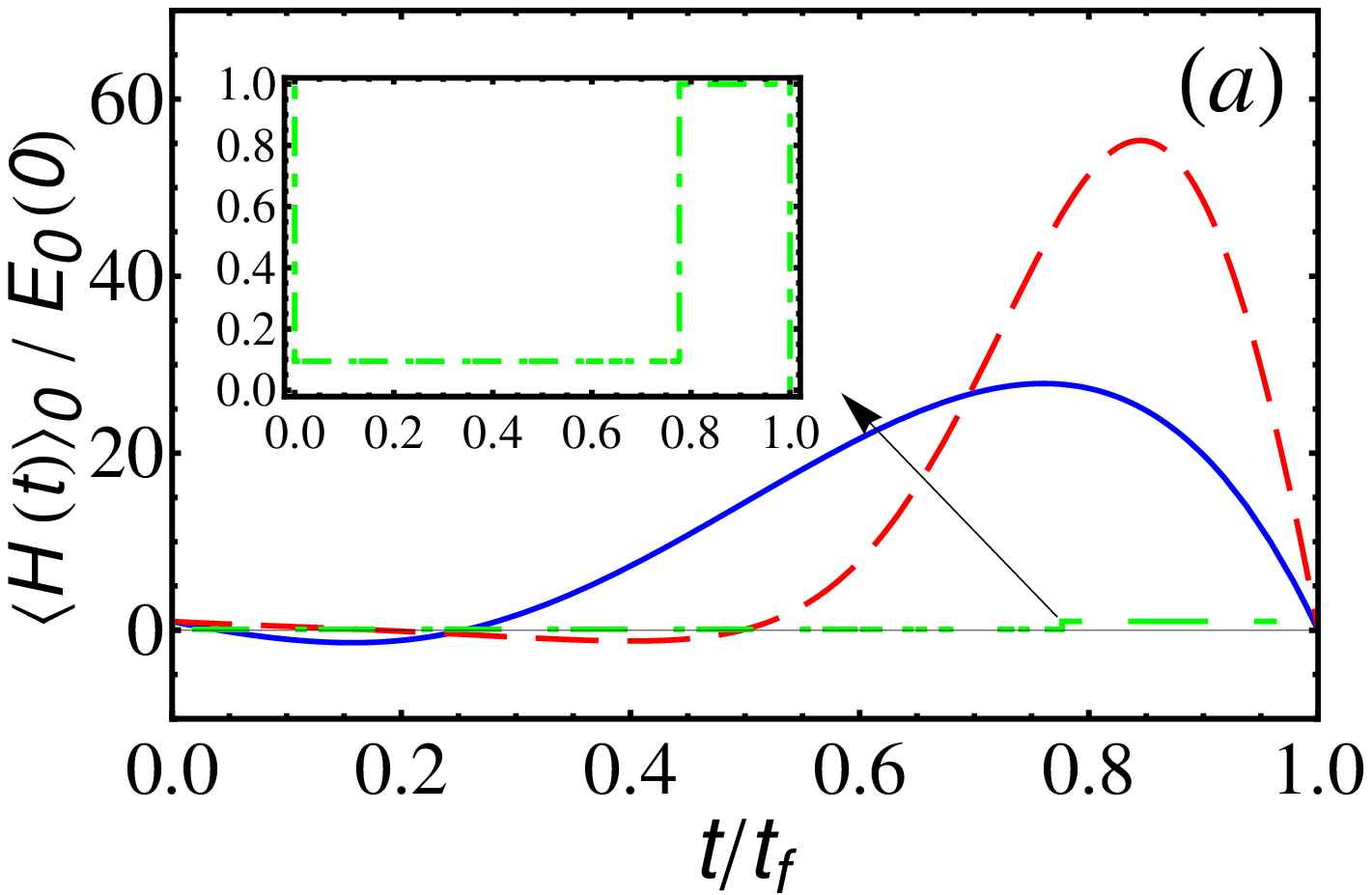}
%
%
\includegraphics[width=0.46\linewidth]{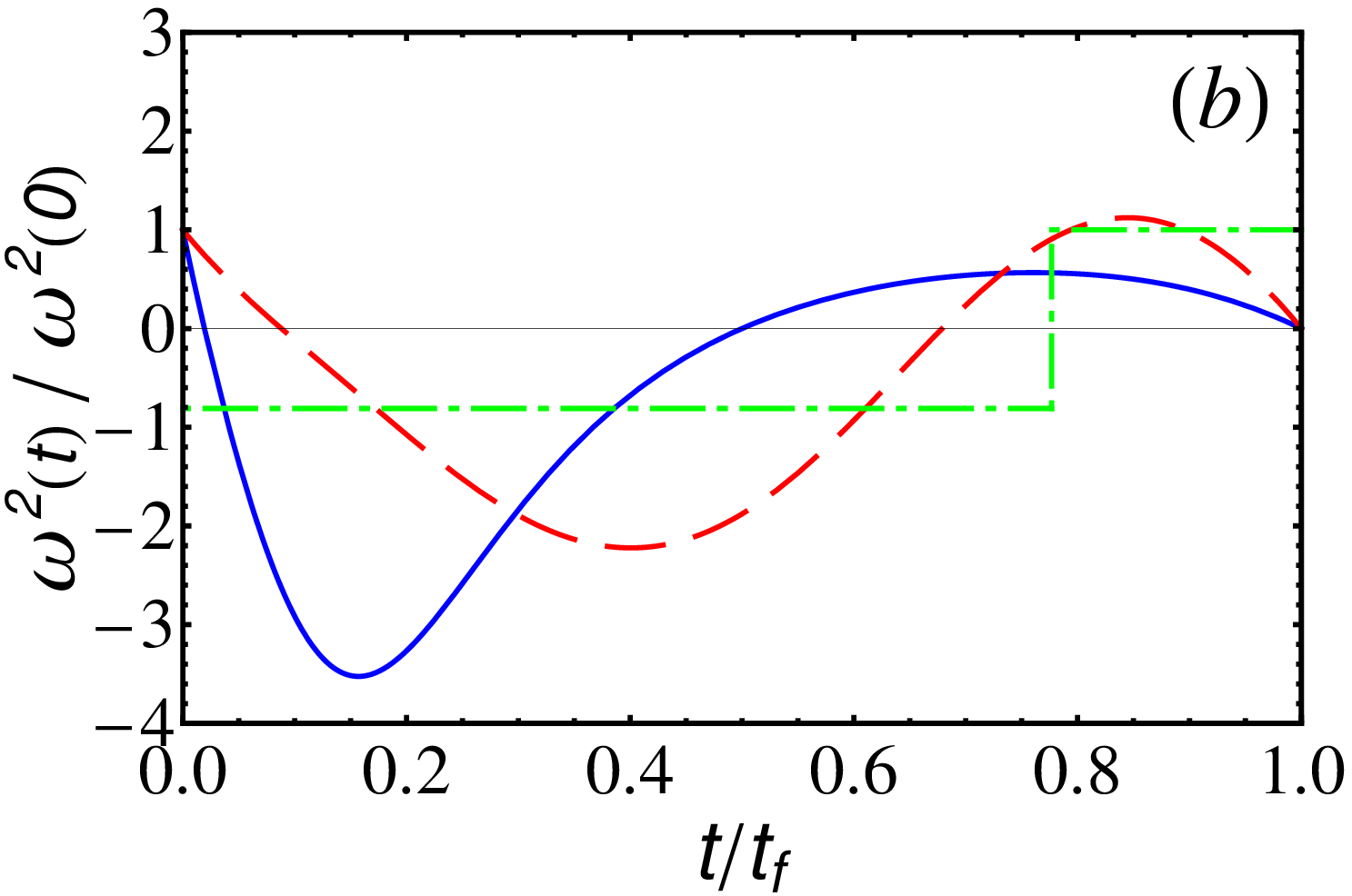}
\end{center}
\caption{\label{fig3}(color online). Cooling in $t_f=2 \ms$. (a) Average energies of the ground expanding mode for $b$ taken as a polynomial (solid),
as an exponential of a polynomial (dashed), and for a piecewise constant frequency ``bang-bang'' process (dot-dashed) with $\omega_1=i 0.9 \omega_0$ and $\omega_2=\omega_0$.  
Other parameters as in Fig. \ref{fig1}.
(b) The squared frequencies $\omega(t)^2$.}
\end{figure}

The appearance of transient energies below the final one, see e.g. the solid line in Fig. \ref{fig3}a
near $t/t_f=0.15$, may be misleading. It is a consequence of the repulsive regime and should not be
interpreted as useful cooling in a time shorter than $t_f$. 
Since the ``trap''  is actually a repeller the 
kinetic energy would grow without bound if the potential were kept
frozen at the time when the energy is minimal. 
Similarly, if the potential were suddenly changed into its final form, $V(t_f)$, the total energy would be higher than the 
adiabatic energy, i.e., the one for a population-preserving process.     

%
%
Fig. \ref{fig3} illustrates 
that a given cooling objective may be
attained in less time than the minimal time required by real-frequency bang-bang trajectories, optimal among real-frequency trajectories \cite{K09}.
For the three-jump
``trajectory'' \cite{K09},
\begin{eqnarray}
\omega(t)= \left\{
\begin{array}{ll}
\omega_0 & (t=0) \\
\omega_1 & (0 < t < \tau_1)\\
\omega_2 & (\tau_1 < t < \tau_1 + \tau_2)\\
\omega_f & (t = t_f=\tau_1 + \tau_2)
\end{array}
\right.
\end{eqnarray}
the smaller $\omega_1$ and the larger $\omega_2$ are, the faster the cooling 
is.
Thus the fastest process to reach the target state corresponds to
the limit of $\omega_1 \rightarrow 0$ and $\omega_2 \rightarrow \infty$ \cite{K09} with   
\beq\label{tmin}
t^{min}_f = \frac{\sqrt{1- \omega_f/ \omega_0}}{\sqrt{\omega_f \omega_0}}.
\eeq
These results are based on optimal control theory, initial and final thermal states, and the constraint $\omega_{1,2}>0$. Clearly, relaxing the positivity condition for the intermediate frequencies, makes faster processes with $t_f<t^{min}_f$ possible, which, moreover, involve only finite frequencies. 
Since Eq. (\ref{tmin}) has been used to justify a finite time version of the third principle (if $\omega_f\to0$, $t^{min}_f\to\infty$ as $\omega_f^{-1/2}$) and maximal cooling rates, the present findings call for a revision of these conclusions.  
A bang-bang example is shown in Fig. 3
(dot-dashed lines), for   
$t_f=2 \ms$, much shorter than the time $t^{min}_f\approx 
6 \ms$ corresponding to the initial and final frequencies chosen.  
$\omega_1=i \omega_I$ is imaginary, and the corresponding 
$b_1(t)$ solving the Ermakov equation 
with initial conditions $b_1(0)=1$ and $\dot{b_1}(0)=1$,
takes the form
$
b_1(t)= [{1 + \frac{\omega^2_0+\omega^2_I}{\omega^2_I} \sinh^2{(\omega_I t )}}]^{1/2}.
$
In the second segment we assume $\omega_2$ real which gives, with the final conditions $b_2(t_f)=\gamma$
and $\dot{b_2}(t_f)=0$,
$b_2(t)= \{{\gamma^2 + \left(\frac{\omega^2_0}{\omega^2_2 \gamma^2}- \gamma^2 \right) \sin^2{[\omega_2 (t-t_f)]}}\}^{1/2}$. 
The matching conditions $b_1(\tau_1)=b_2(\tau_1)$ and $\dot{b_1}(\tau_1)=\dot{b_2}(\tau_1)$
are then solved for $\tau_1$ and $t_f$, 
see Fig. 3 and its caption for details. Of course the discontinuous
jumps in this type of  trajectory call into question its realizability.   
Fig. 3 also shows two smooth trajectories 
for $t_f= 2 \ms$ corresponding to two different ansatz for $b$ (polynomial and exponential of a polynomial). 
The resulting rates of change of $\omega(t)$ are feasible with present
technology. Indeed  the intensity of a dipole trap (the frequency scales as the
square root of the intensity) can be changed by three or four orders
of magnitude  in $100$ ns using acousto-optics or electro-optics modulators.
To monitor the sign of the square frequencies, one can
superimpose two dipole beams locked respectively on the blue and red
side of the line.
By alternating them rapidly with a control of their relative
intensity, one can shape the square
frequencies and their signs at will. Alternatively, one can combine
magnetic and dipole traps.  
%
\begin{figure}[t]
\begin{center}
%
\includegraphics[width=0.485\linewidth]{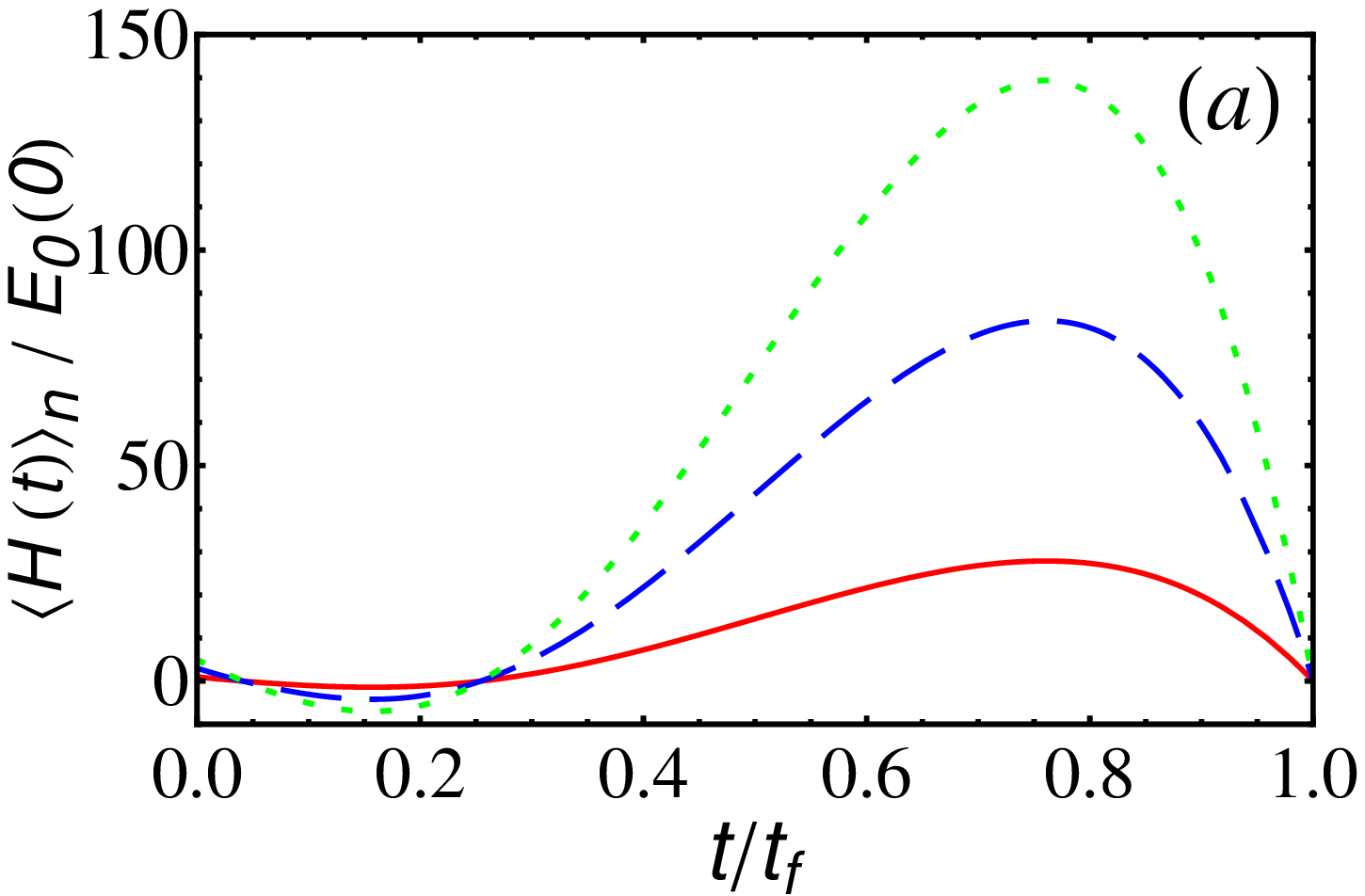}
%
%
\hspace{.2cm}\includegraphics[width=0.48\linewidth]{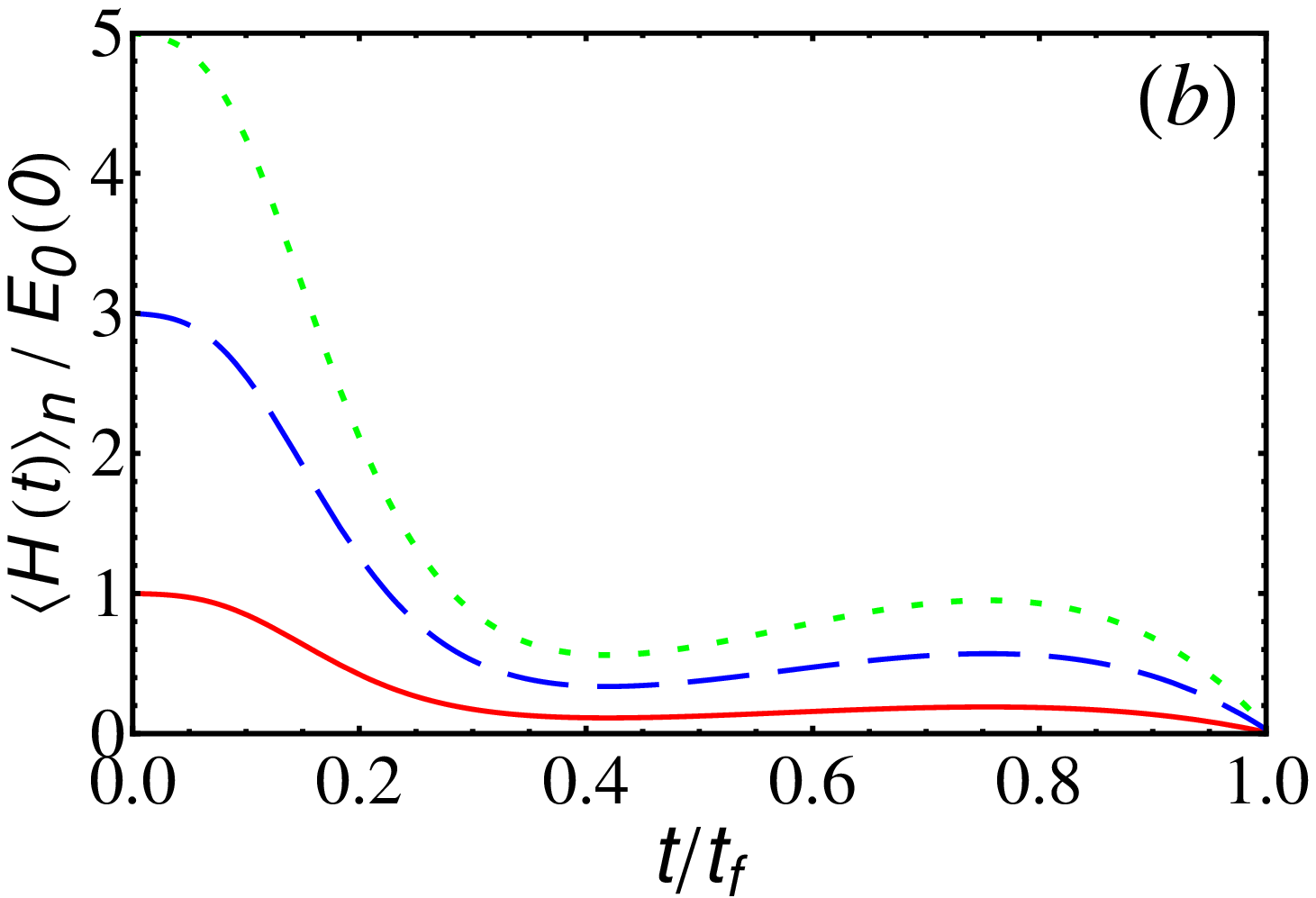}
\end{center}
\caption{\label{fig4} (color online). Average energies for expanding modes $n=1$ (solid), $n=2$ (dashed), and $n=3$ (dotted). (a) $t_f = 2 \ms$; (b) $t_f = 25 \ms$.
Other parameters as in Fig. \ref{fig1} (polynomial $b$).}
\end{figure}

Finally, while the initial (and final) state in Figs. 2 and 3 is the ground state, Fig. 4 illustrates that the same $\omega(t)$ trajectories work as well for arbitrary excited states.  
This also means that fast frictionless cooling is directly 
applicable to arbitrary superpositions or mixed states, as well as to simple many body systems such as a polarized Fermi
gas, or its (symmetrized) bosonic counterpart,
the Tonks-Girardeau gas.
Similar techniques may also be applicable for weakly interacting bosonic systems, the control of soliton dynamics of Bose-Einstein condensates \cite{Salomon,Kono}, and to manipulate the transport of ultracold atoms \cite{David2}.

We have in summary described a method for fast, frictionless cooling 
in a harmonic trap based on a modulation of the trap frequency $\omega(t)$  
that includes in general time intervals in which the potential becomes expulsive.
It is applicable to arbitrary initial states, not necessarily in equilibrium.  The algorithm to design $\omega(t)$ depends on flexible ansatz functions that can be modified or made more and more complex, by adding parameters, to satisfy further requirements. This could be used, e.g., to minimize the maximal frequencies along the trajectory.       

We are grateful to M. Berry and L. S. Schulman for discussions, and  
to the Max Planck Institute for Complex Systems at Dresden for hospitality.
We acknowledge funding by Projects No. GIU07/40, FIS2006-10268-C03-01,
60806041, 08QA14030, 2007CG52, S30105, ANR-09-BLAN-0134-01, Juan de la Cierva Program, EU Integrated
Project QAP, EPSRC project EP/E058256,
and the German Research Foundation (DFG).

\end{document}